\documentclass[aps,prl,reprint,showpacs,superscriptaddress]{revtex4-1}% Physical Review B
\usepackage{graphicx}% Include figure files
\usepackage{latexsym}
\usepackage[T1]{fontenc}
\usepackage{amsmath}
\usepackage{endnotes}
\usepackage{lmodern}
\input{epsf}

\begin{document}

\preprint{APS/123-QED}

\date{\today}

\title{Nematic susceptibility of hole-doped and electron-doped BaFe$_2$As$_2$ iron-based superconductors from shear modulus measurements}

\author{A. E. Böhmer}
\email{anna.boehmer@kit.edu}
\affiliation{Institut für Festkörperphysik, Karlsruhe Institute of Technology, 76021 Karlsruhe, Germany}
\affiliation{Fakultät für Physik, Karlsruhe Institute of Technology,  76131 Karlsruhe, Germany}

\author{P. Burger}
%\email{philipp.burger@ifp.fzk.de}
\affiliation{Institut für Festkörperphysik, Karlsruhe Institute of Technology, 76021 Karlsruhe, Germany}
\affiliation{Fakultät für Physik, Karlsruhe Institute of Technology,  76131 Karlsruhe, Germany}

\author{F. Hardy} 
%\email{frederic.hardy@kit.edu}
\affiliation{Institut für Festkörperphysik, Karlsruhe Institute of Technology, 76021 Karlsruhe, Germany}

\author{T. Wolf}
%\email{thomas.wolf2@kit.edu}
\affiliation{Institut für Festkörperphysik, Karlsruhe Institute of Technology, 76021 Karlsruhe, Germany}

\author{P. Schweiss}
%\email{peter.schweiss@kit.edu}
\affiliation{Institut für Festkörperphysik, Karlsruhe Institute of Technology, 76021 Karlsruhe, Germany}

\author{R. Fromknecht}
%\email{rainer.fromknecht@kit.edu}
\affiliation{Institut für Festkörperphysik, Karlsruhe Institute of Technology, 76021 Karlsruhe, Germany}

\author{M. Reinecker}
\affiliation{Faculty of Physics, University of Vienna, Boltzmanngasse 5, Vienna A-1090, Austria}

\author{W. Schranz}
\affiliation{Faculty of Physics, University of Vienna, Boltzmanngasse 5, Vienna A-1090, Austria}

\author{C. Meingast}
\email{christoph.meingast@kit.edu}
\affiliation{Institut für Festkörperphysik, Karlsruhe Institute of Technology, 76021 Karlsruhe, Germany}

\begin{abstract}

The nematic susceptibility, $\chi_\varphi$, of hole-doped Ba$_{1-x}$K$_x$Fe$_2$As$_2$ and electron-doped Ba(Fe$_{1-x}$Co$_x$)$_2$As$_2$ iron-based superconductors is obtained from measurements of the elastic shear modulus using a three-point bending setup in a capacitance dilatometer. Nematic fluctuations, although weakened by doping, extend over the whole superconducting dome in both systems, suggesting their close tie to superconductivity. Evidence for quantum critical behavior of $\chi_\varphi$ is, surprisingly, only found for Ba(Fe$_{1-x}$Co$_x$)$_2$As$_2$ and not for Ba$_{1-x}$K$_x$Fe$_2$As$_2$ - the system with the higher maximal $T_c$ value.
%Enhanced nematic fluctuations appear to be closely tied to superconductivity. However, the nematic susceptibility of the system with the higher $T_c$, (Ba,K)Fe$_2$As$_2$, shows no quantum criticality and the weaker response to superconductivity, suggesting that this relation might not be fundamental.
% This susceptibility, which drives the structural transition, is enhanced over most of the wide superconducting dome of both Co-Ba122 and K-Ba122, providing evidence for the persistence of fluctuations of $\varphi$ up to at least 80\% hole doping. These fluctuations become quantum critical only in the electron-doped compound suggesting that quantum criticality is not necessary for high superconducting transition temperatures. 
  %The reliability of the technique is confirmed by comparing results on Ba(Fe,Co)$_2$As$_2$ with previous ultrasound studies [Yoshizawa \textit{et al.}, J. Phys. Soc. Jpn. 81, 024604]. 
\end{abstract}

% Mechanical and acoustical properties, elasticity, and ultrasonic attenuation in superconductors, Quantum critical phenomena in superconductors, effects of pressure in superconductors, pnictides

\maketitle

%\section{Introduction}\label{sec:intro}

As in many other unconventional superconductors, superconductivity in iron-based superconductors arises close to the point where an antiferromagnetic (AFM) spin-density-wave transition (SDW) is suppressed\cite{Norman2011, Stewart2011}. A particular feature of the iron-based materials is that the SDW breaks the four-fold rotational symmetry of the lattice which is accompanied by an orthorhombic distortion\cite{Stewart2011}. The fact that the structural transition surprisingly precedes the SDW transition in some systems has led to a strong debate about the driving force of the structural phase transition; \textit{e.g.}, orbital\cite{Lee2009, Lv2010, Kontani2011} or spin-nematic\cite{Nandi2010, Fernandes2012} degrees of freedom have been proposed. Importantly, orbital or spin fluctuations are also likely candidates for the superconducting pairing glue\cite{Mazin2008,Kontani2010}.  Whereas magnetic fluctuations are believed to mediate $s_\pm$-superconductivity\cite{Mazin2008,Kuroki2008,Mazin2009}, orbital fluctuations are thought to lead to $s_{++}$-superconductivity\cite{Kontani2010,Kontani2011}. 
%In the context of the structural transition in the Fe-based materials, the physics of electronic nematic phases is of strong current interest\cite{Chu2010,Yi2011,Kasahara2012,Chu2012,Fernandes2012}. In these intriguing phases, which occur also in Sr$_3$Ru$_2$O$_7$ and cuprate superconductors\cite{Vojta2009}, the electron fluid spontaneously breaks the rotational symmetry\cite{Fradkin2010}.
The structural transition in the Fe-based materials is of strong current interest\cite{Chu2010,Yi2011,Kasahara2012,Chu2012,Fernandes2012,Gallais2013}, because it is believed to be of an electronic nematic type, \textit{i.e.} a transition in which a ``nematic'' order parameter of electronic (\textit{e.g.} spin or orbital) origin causes a spontaneous breaking of the four-fold lattice symmetry. Such nematic transitions are believed to occur also in Sr$_3$Ru$_2$O$_7$ and the cuprate superconductors\cite{Fradkin2010,Vojta2009}.

Measurements of the elastic shear modulus $C_{66}$, which is the soft mode of the structural transition, provide a powerful tool for studying the susceptibility and associated fluctuations of this nematic order parameter\cite{Fernandes2010,Goto2011,Yoshizawa2012}. In particular, the softening of $C_{66}$ was studied intensively in Ba(Fe$_{1-x}$Co$_x$)$_2$As$_2$ (Co-Ba122) using ultrasound, and strong evidence for a quantum critical point (QCP) was obtained near optimal doping\cite{Yoshizawa2012}. Further, the response of $C_{66}$ below the superconducting transition at $T_c$ provides important information about the coupling of these fluctuations to superconductivity, which has been interpreted both in the spin-nematic and the orbital scenario\cite{Fernandes2010, Yoshizawa2012}.
%Here, the QCP was interpreted as being a structural QCP due to orbital physics\cite{Yoshizawa2012}, however spin-nematic fluctuations may also lead to similar effects\cite{Fernandes2010}. 
Up to the present, shear-modulus data have only been reported for the electron-doped Co-Ba122 system from ultrasound measurements. It is clearly of interest to establish whether the nematic susceptibility of other Fe-based superconductors shows similar behavior. 
%Up to the present, data on the nematic susceptibility is only available for the Co-Ba122 system, and it is clearly of interest to establish whether the nematic susceptibility of other iron-based superconductors shows similar behavior. litshear-modulus data are only available for the Co-Ba122 system from ultrasound measurements, and it is clearly of interest to obtain information on nematic fluctuations in other iron-based superconductors. 

\begin{figure*}
\includegraphics[width=17cm]{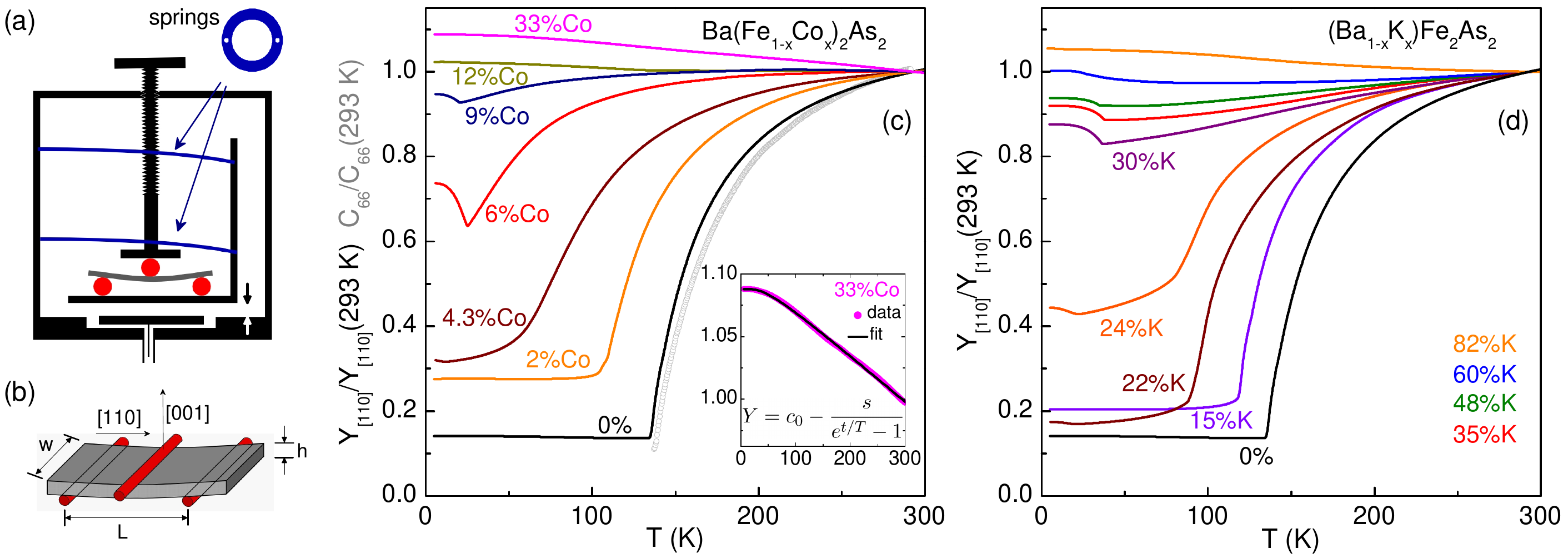}
\caption{(color online) (a) Schematic drawing of our three-point-bending setup in the capacitance dilatometer. The sample (grey) is supported by three wires (red) and the capacitor gap is indicated by small arrows. (b) Definition of sample dimensions and orientation relative to the wire supports. Young's modulus $Y_{[110]}$ of (c) Co-Ba122 and (d) K-Ba122, normalized at room temperature. For comparison the temperature dependence of the $C_{66}$ mode of pure BaFe$_2$As$_2$ from Ref. \onlinecite{Yoshizawa2012} (grey circles) is shown in (c). The inset in (c) shows a fit of the 33\%Co-data to the Varshni formula used as the phonon background.}
\label{fig:1}
\end{figure*}

In this Letter we present an extensive study of the nematic susceptibility derived from shear modulus data of both hole-doped Ba$_{1-x}$K$_x$Fe$_2$As$_2$ (K-Ba122) and electron-doped Co-Ba122 covering a wide doping range. We find that nematic fluctuations are enhanced over the whole superconducting dome of both systems, suggesting their close tie to superconductivity. Surprisingly, evidence for quantum critical behavior of the nematic susceptibility is only found for Co-Ba122, and not for K-Ba122 - the system with the higher maximal superconducting transition temperature $T_{c,max}$.
%This raises the question as to which degree they are involved in superconducting pairing. 
%The nematic susceptibility of a Landau theory is found to be enhanced over large parts of the superconducting domes of both Co-Ba122 and K-Ba122, however, only the electron-doped system exhibits quantum criticality. 
%This suggests that whereas nematic fluctuations are possibly related to superconductivity, quantum criticality is not a necessary ingredient for high transition temperatures. 

%, which could mediate superconducting pairing, persist on both sides,  On the other hand, quantum criticality, while confirmed for electron-doped Co-Ba122 is found to be absent in hole-doped K-Ba122, suggesting that quantum critical fluctuations are not necessary for realizing high-temperature superconductivity.

Self-flux grown single crystals of Co-Ba122 and K-Ba122 were cut to dimensions of $L\times w\times h\sim(2-3\times1\times0.1)\textnormal{ mm}^3$ (see Fig. \ref{fig:1}(b)). We gain access to their shear modulus $C_{66}$ via a measurement of the Young's modulus (\textit{i. e.} the modulus of elasticity for uniaxial tension) along the tetragonal [110] direction $Y_{[110]}$ using the novel technique of a three-point bending setup in a capacitance dilatometer\cite{Meingast1990,noteSM} (see Fig. \ref{fig:1} (a,b)). If $C_{66}$ is small, it is expected to dominate $Y_{[110]}$\cite{Kityk1996,noteSM}. This certainly holds close the the structural transition, but, as we demonstrate on the Co-Ba122 system, our method yields very similar results as the ultrasound data over the whole doping range\cite{Yoshizawa2012}, showing that the temperature dependence of $C_{66}$ may be reliably obtained also by three-point bending. Fig. \ref{fig:1}(c,d) shows the Young's modulus $Y_{[110]}/Y_{[110]}(293\textnormal{ K})$ of Co-Ba122 and K-Ba122, which was normalized at room temperature because of uncertainties in the geometrical parameters $L$, $w$ and $h$. Importantly, our data on Co-Ba122 agree very well with the $C_{66}$ results of Ref. \onlinecite{Yoshizawa2012} over the whole doping range; for a direct comparison we have also plotted $C_{66}$ data for $x=0$ in Fig. \ref{fig:1}(c). 

Both characteristics of the electron-doped system, namely the softening on approaching the structural transition at $T_s$ from above and the hardening below $T_c$, are observed also in hole-doped K-Ba122.  However, the softening and subsequent hardening, at \textit{e.g.} optimal doping, is much more pronounced for Co- than for K-doping, which will be discussed in more detail later. 

In general, the elastic constants associated with the soft mode are expected to go to zero at a second-order structural phase transition and to harden below\cite{Salje1990}. Surprisingly, this is not the case here and $Y_{[110]}$, even though it decreases by 50-85\%, never reaches zero, a behavior which is not fully understood and also observed in ultrasound data\cite{Yoshizawa2012}. Also, we find that $T_s$ manifests itself as a kink in $Y_{[110]}(T)$ and that $Y_{[110]}$ stays soft, or even grows softer below $T_s$, contrary to the general expectation. This effect presumably arises from ``superelastic'' behavior\cite{Schranz2012II,Schranz2012}, \textit{i.e.} from the motion of boundaries between structural twins that are formed in the orthorhombic phase.

As argued previously\cite{Chu2012,Yoshizawa2012II}, the structural transition in Ba122 is most likely driven by an electronic order parameter $\varphi$ via bilinear coupling to the orthorhombic strain $\varepsilon_6=(a-b)/(a+b)$. ($a$ and $b$ are the in-plane lattice constants in the orthorhombic phase.) In this case the Landau expansion of the free energy is given by
\begin{equation}
F=\frac{1}{2}\left(\chi_\varphi\right)^{-1}\varphi^2+\frac{B}{4}\varphi^4+\frac{C_{66,0}}{2}\varepsilon_6^2-\lambda\varphi\varepsilon_6,
\end{equation}
where $\lambda$ is the electron-lattice coupling-constant and $C_{66,0}$ the bare elastic constant, which has no strong temperature dependence and $B$ is the quartic coefficient of the Landau expansion. 
%Bilinear coupling is allowed if $\varphi$ and $\varepsilon_6$ transform under the same irreducible representation. Thus, $\varphi$ also breaks the four-fold rotational symmetry of the lattice and we will therefore refer to it as the ``nematic'' order parameter. 
Bilinear coupling is allowed because $\varphi$ and $\varepsilon_6$ both break the same four-fold rotational symmetry. We therefore refer to $\varphi$ as the nematic order parameter with $\chi_\varphi$ the associated nematic susceptibility. $\varphi$ may represent \textit{e.g.} the spin-nematic or the orbital order parameter, however, the present thermodynamic treatment cannot distinguish between these scenarios. $C_{66}$ is renormalized due to the coupling $\lambda$\cite{Salje1990, Fernandes2010, Cano2010}, and is given by
%$C_{66,0}$ is renormalized in this approach because the system can relax $\varphi$ under external stress and the effective shear modulus is given by\cite{Cano2010, Salje1990}
\begin{equation}
{C_{66}}=C_{66,0}-\lambda^2\chi_\varphi. \label{eq:5}
\end{equation}
At a mean field level, the temperature dependence of $\chi_\varphi$ is given by $\chi_\varphi=\left[A(T-T_0)\right]^{-1}$, reflecting that $\chi_\varphi$ diverges at the ``bare'' transition temperature $T_0$, \textit{i.e.} the nematic ordering temperature in the absence of electron-lattice coupling. Due to the coupling $\lambda$, however, the ordering of $\varphi$ and the associated structural distortion takes place at $T_s^{CW}=T_0+\lambda^2/AC_{66,0}$, the temperature at which $\chi_\varphi$ reaches the critical value $C_{66,0}/\lambda^2$. $C_{66}$ in turn follows the following modified Curie-Weiss law
\begin{equation}
\frac{C_{66}}{C_{66,0}}=\frac{T-T_s^{CW}}{T-T_0}.\label{eq:7}
\end{equation}
The difference $T_s^{CW}-T_0=\lambda^2/AC_{66,0}$ (the ``Jahn-Teller energy'' of Refs. \onlinecite{Goto2011,Yoshizawa2012}) is an energy scale characteristic of the electron-lattice coupling. 

\begin{figure}
\includegraphics[width=8.6cm]{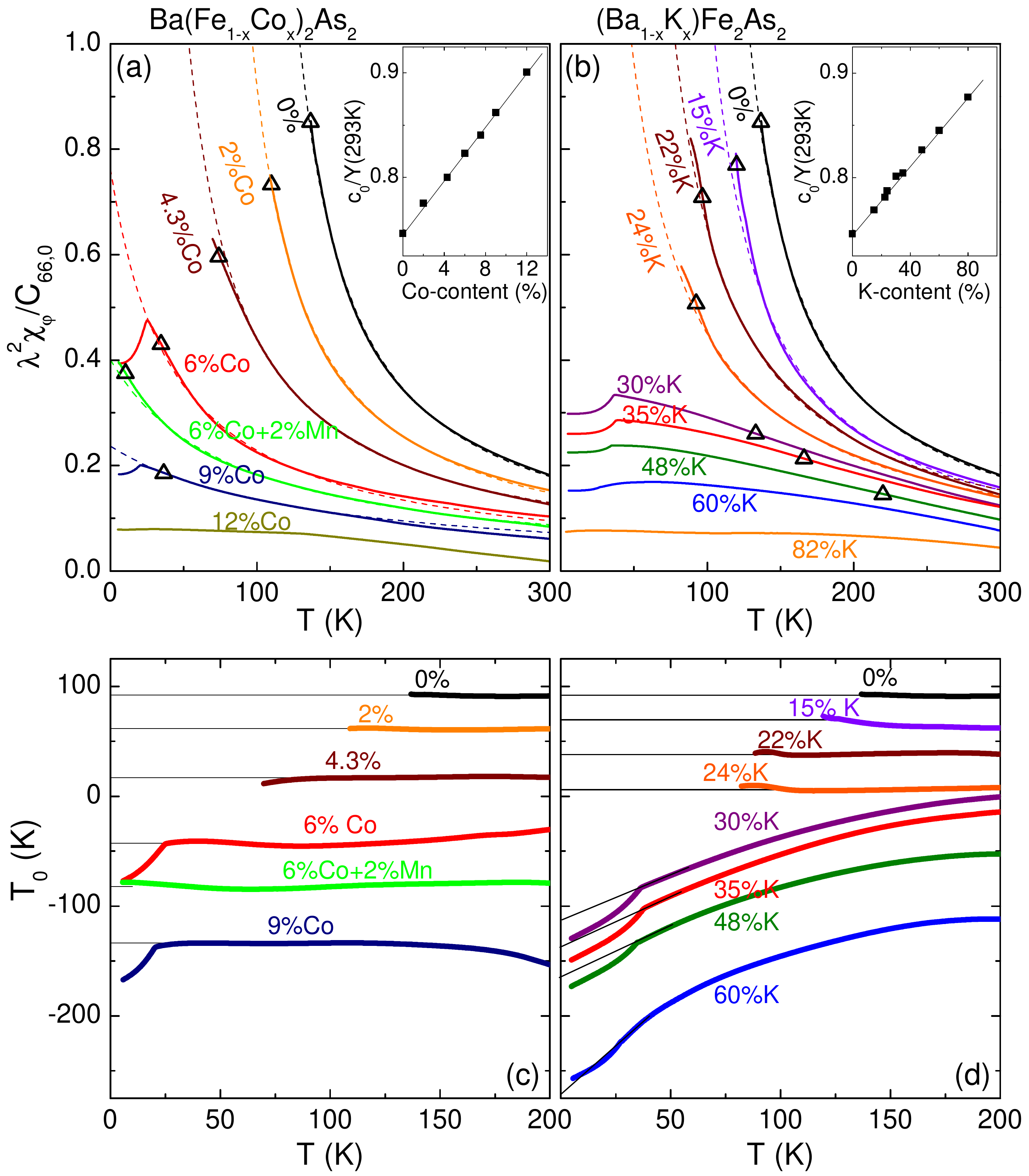}
\caption{(color online). Nematic susceptibility, $\chi_\varphi$ in units of $\lambda^2/C_{66,0}$, obtained from the critical part of the Young's modulus of (a) Co-Ba122 and (b) K-Ba122. Dashed lines correspond to a Curie-Weiss fit using equation \ref{eq:7}. Insets show the parameter $c_0/Y_{[110]}(293\textnormal{ K})$ used for separating the critical part of $Y_{[110]}$ from the background (see text for details). Triangles mark the inflection point of $\chi_\varphi(T)$, $T^*$, which is a lower temperature limit for its Curie-Weiss-like divergence. (c),(d) Temperature-dependent Curie-Weiss temperature $T_0(T)=1-1/A\chi_\varphi$ of the nematic susceptibility for (c) Co-Ba122 and (d) K-Ba122. Linear extrapolations beyond the structural or superconducting transition are shown as thin black lines.}
\label{fig:2}
\end{figure}

In the following we extract the nematic susceptibility $\chi_\varphi$ from our data using the above Landau theory. We use the approximation $C_{66}/C_{66,0}\approx Y_{[110]}/Y_0$, where $Y_0$ is the non-critical background. For $Y_0$, we use 33\%Co-Ba122 data\cite{Meingast2012}, the temperature-dependence of which can be very well described by the empirical Varshni-formula\cite{Varshni1970} $Y_0=c_0-s/(\exp(t/T)-1)$ with $s/c_0=0.0421$ and $t=123.6$ K (see inset in Fig. \ref{fig:1}(c)). $c_0/Y(293\textnormal{ K})$ remains a free parameter, because we lack absolute values of $Y$. The values of $c_0/Y(293\textnormal{ K})$ were adjusted in order to obtain good agreement with equation \ref{eq:7} for the underdoped samples and then linearly extrapolated to higher doping levels (see insets of Fig. \ref{fig:2}). Making use of equation \ref{eq:5}, the normalized nematic susceptibility $\frac{\lambda^2\chi_\varphi}{C_{66,0}}$ can thus be obtained from our data (Fig. \ref{fig:2} (a),(b)). 
%$\chi_\varphi$ is the susceptibility which ultimately drives the structural transition. It would diverge at $T_0$ which is the transition temperature of $\varphi$ in the absence of coupling to the lattice. In this dimensionless form, the structural transition should occur when $\lambda^2\chi_\varphi/C_{66,0}=1$ (i.e. $C_{66}=0$). As can be seen in Fig. \ref{fig:2}(a) and \ref{fig:2}(b), $\lambda^2\chi_\varphi/C_{66,0}$ closely follows a Curie-Weiss behavior, however it never reaches $1$, which is not fully understood.
%We first discuss the doping dependence of the magnitude and then the detailed temperature dependence of $\chi_\varphi$. 

\begin{figure}
\includegraphics[width=8.6cm]{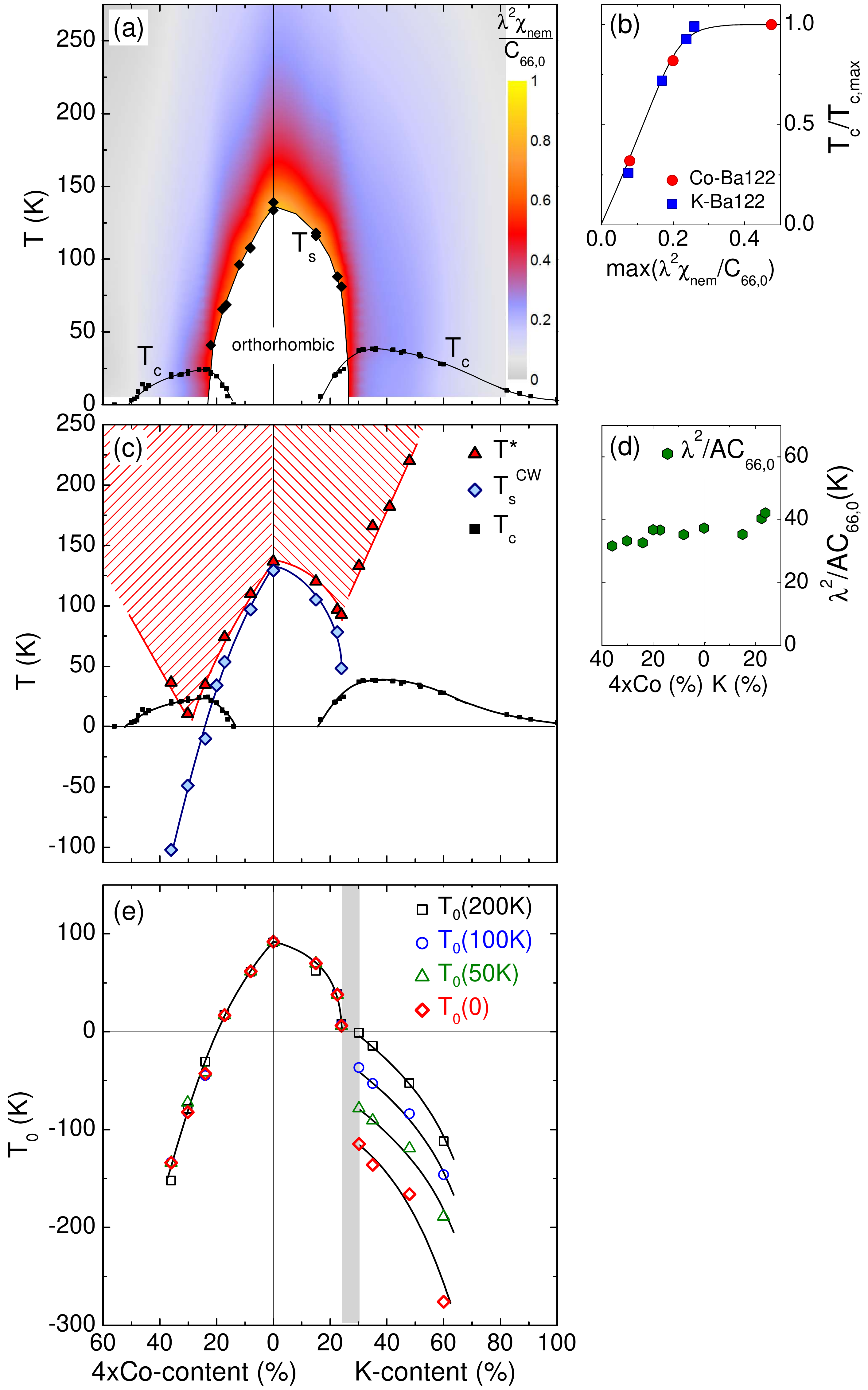}
\caption{(color online). (a) Temperature and doping dependence of the magnitude of the normalized nematic susceptibility $\lambda^2\chi_\varphi/C_{66,0}$ for Co-Ba122 and K-Ba122 from Fig. \ref{fig:2}. (b) $T_c/T_{c,max}$ as a function of the maximum value of $\frac{\lambda^2\chi_\varphi}{C_{66,0}}(T)$ for all overdoped samples. (c) Phase diagram showing the doping dependence of $T^*$, $T_s^{CW}$ and $T_c$. The red dashed area corresponds to the ``critical'' region above $T^*$ where $\chi_\varphi(T)$ diverges. (d) Doping dependence of the electron-lattice coupling energy $T_s^{CW}-T_0=\frac{\lambda^2}{AC_{66,0}}$. (e) Curie-Weiss temperature $T_0$ at various temperatures obtained from Fig. \ref{fig:2}(c),(d). An abrupt change from Curie-Weiss to non-Curie-Weiss behavior occurs in the shaded region. Lines are a guide to the eye.}
\label{fig:3}
\end{figure}

Fig. \ref{fig:3}(a) shows the magnitude of the nematic susceptibility $\chi_\varphi$ (in units of $\lambda^2/C_{66,0}$) as a color-coded map in the composition-temperature phase diagram of Co-Ba122 and K-Ba122. 
$\chi_\varphi$ is significantly enhanced in a broad band around the structurally distorted phase, as expected. It is largest for the undoped compound right above $T_s$ and decreases smoothly with both electron and hole doping. This is at variance with Ref.\cite{Chu2012}, where a maximum of the nematic susceptibility around optimal doping has been reported.
Importantly, the enhancement of $\chi_\varphi$ extends over most of the superconducting domes of both systems, even when no structural phase transition occurs. In fact, there appears to be a universal relationship between the ratio $T_c/T_{c,max}$ [$T_{c,max}=25$ K (38 K) for Co-Ba122 (K-Ba122)] and the maximum value of $\chi_\varphi(T)$ for the overdoped samples of both systems (Fig. \ref{fig:3}(b)). If the nematic susceptibility is taken as a measure of the strength of nematic fluctuations, this suggests that these fluctuations may mediate the superconducting pairing also over the whole, rather wide, superconducting dome of K-Ba122. 
%. This suggests a close tie between the nematic susceptibility and the occurrence of superconductivity for both electron- and hole-doped Ba122 systems.   

%\begin{figure}
%\includegraphics[width=8.6cm]{Fig4v2.pdf}
%\caption{(color online). }
%\label{fig:4}
%\end{figure}

%The temperature dependence of $\chi_\varphi$ however shows that nematic fluctuations become quantum critical only in electron-doped Co-Ba122 and not in hole-doped K-Ba122.
In Fig. \ref{fig:2} we also show the results from a Curie-Weiss analysis of the nematic susceptibility. The dashed lines in Fig. \ref{fig:2}(a),(b) show a fit of the data to equation \ref{eq:7}, which takes the form $\frac{\lambda^2\chi_\varphi}{C_{66,0}}(T)=\frac{T_s^{CW}-T_0}{T-T_0}$. The resulting parameters are plotted in Fig. \ref{fig:3}(c),(d). The coupling energy $T_s^{CW}-T_0=\lambda^2/AC_{66,0}\sim30-40$ K is practically doping independent. In the electron-doped system, $T_s^{CW}$ changes smoothly from positive to negative values, a behavior which has been associated with a QCP at optimal doping\cite{Yoshizawa2012}. Note that the hole-doped system with K-content $\geq30\%$ cannot be described successfully by the simple Curie-Weiss law.  

In order to study how hole-doped K-Ba122 differs from Co-Ba122, we define a lower temperature limit for the breakdown of the Curie-Weiss law by the inflection point of $\chi_\varphi(T)$, $T^*$. $T^*$ has been marked in Fig. \ref{fig:2}(a),(b) by open triangles and is also plotted in Fig. \ref{fig:3}(c). As a function of doping, $T^*$ first decreases, closely following $T_s$, and then increases upon further doping. The region of possible critical softening above $T^*$ extents to near zero temperatures only in the electron-doped system, in agreement with a quantum critical scenario. Note that, in order to ``look beneath'' the superconducting dome, we used a non-superconducting 6\%Co+2\%Mn-codoped sample, because Mn substitution strongly suppresses $T_c$ but affects $T_s$ only slightly\cite{noteCoMnBa122}. In contrast, the values of $T^*$ for the hole-doped system clearly do not go below $\sim75$ K, which is incompatible with a QCP in this system. It is curious that optimally doped K-Ba122, the sample with the highest $T_c$, does not show Curie-Weiss-like critical softening. Note that also the hardening below $T_c$ is significantly less pronounced in optimally doped K-Ba122 than in optimally doped Co-Ba122. For a detailed description of the anomalies at and below $T_c$ see \cite{noteSM}. 

In order to better quantify the deviations from the Curie-Weiss behavior, we ascribe them to a temperature dependence of the parameter $T_0$ in equation \ref{eq:7}. $T_0(T)$ can be obtained from the data in Fig.\ref{fig:2}(a),(b) assuming that no other parameter is temperature dependent, and that $\lambda^2/AC_{66,0}=40$ K for K-Ba122 with $\geq30\%$ K content. Fig.\ref{fig:2}(c),(d) shows that $T_0(T)$ is temperature independent for Co contents up to 9\% and K contents up to 24\% confirming that the nematic susceptibility obeys the Curie-Weiss law for these compounds. In contrast, $T_0(T)$ is strongly temperature dependent for higher K contents. This contrasting behavior is made even more evident in Fig. \ref{fig:3}(e) in which $T_0$ at different temperatures is plotted as a function of doping. A linear extrapolation is used for temperatures which lie below the structural or superconducting phase transitions. There is a clear abrupt change in behavior between 24\% and 30\% K content, above which $T_0$ is no longer constant and decreases with decreasing temperature below 200 K. Especially at low temperatures, a step-like anomaly of $T_0$ as a function of doping occurs, which we associate with a first order transition between orthorhombic/magnetic and tetragonal ground states on increasing the K content.
The decrease of $T_0$ with decreasing temperature shows that the tendency towards nematic ordering is weakened at lower temperature. Interestingly, a change of the topology of the Fermi-surface from strongly nesting circular hole pockets to 'propeller'-shaped hole pockets with expected weaker nesting has been found to occur on lowering the temperature in optimally doped K-Ba122\cite{Evtushinsky2011}. Such a change of topology of the Fermi-surface could explain why $T_0$ becomes temperature dependent. 

In conclusion, we have shown that the nematic susceptibility is enhanced over the whole superconducting regions in both K- and Co-doped Ba122, suggesting that fluctuations associated with the electronic nematic order possibly play a crucial role in superconducting pairing. On the other hand, nematic fluctuations exhibit quantum critical-like behavior only in Co-Ba122 and not in K-Ba122, which has the higher $T_c$. This naturally raises the question whether quantum criticality is really important for obtaining high $T_c$ and/or if nematic fluctuations are indeed directly involved in the pairing. From our measurements alone it is impossible to determine whether nematic order is driven by spin or orbital physics. However, a comparison between the $\chi_\varphi$ presented here and the nematic susceptibility in the spin-nematic model (obtained from nuclear magnetic resonance measurements) was carried out for Co-Ba122\cite{Fernandes2013}, providing strong evidence for a magnetically-driven structural transition. Similar tests with K-Ba122 are highly desirable.

Acknowledgements: We thank Andres Cano, Rafael M. Fernandes, Jörg Schmalian, Kees van der Beek, and Bernd Wolf for valuable discussions. This work was funded by the DFG through SPP1458 and via a ``Feasibility Study of Young Scientists'' in the framework of the Exzellenz-Initiative at the KIT. 

%\bibliography{C:/Users/boehmer/Documents/reports/referencespnictidesall}
%\bibliographystyle{apsrev4-1} 


\newpage

\title{Nematic Susceptibility of Hole-Doped and Electron-Doped BaFe$_2$As$_2$ Iron-Based Superconductors from Shear Modulus Measurements}

\author{A. E. Böhmer}
%\email{anna.boehmer@kit.edu}
\affiliation{Institut für Festkörperphysik, Karlsruhe Institute of Technology, 76021 Karlsruhe, Germany}
\affiliation{Fakultät für Physik, Karlsruhe Institute of Technology,  76131 Karlsruhe, Germany}

\author{P. Burger}
%\email{philipp.burger@ifp.fzk.de}
\affiliation{Institut für Festkörperphysik, Karlsruhe Institute of Technology, 76021 Karlsruhe, Germany}
\affiliation{Fakultät für Physik, Karlsruhe Institute of Technology,  76131 Karlsruhe, Germany}

\author{F. Hardy} 
%\email{frederic.hardy@kit.edu}
\affiliation{Institut für Festkörperphysik, Karlsruhe Institute of Technology, 76021 Karlsruhe, Germany}

\author{T. Wolf}
%\email{thomas.wolf2@kit.edu}
\affiliation{Institut für Festkörperphysik, Karlsruhe Institute of Technology, 76021 Karlsruhe, Germany}

\author{P. Schweiss}
%\email{peter.schweiss@kit.edu}
\affiliation{Institut für Festkörperphysik, Karlsruhe Institute of Technology, 76021 Karlsruhe, Germany}

\author{R. Fromknecht}
%\email{rainer.fromknecht@kit.edu}
\affiliation{Institut für Festkörperphysik, Karlsruhe Institute of Technology, 76021 Karlsruhe, Germany}

\author{M. Reinecker}
\affiliation{Faculty of Physics, University of Vienna, Boltzmanngasse 5, Vienna A-1090, Austria}

\author{W. Schranz}
\affiliation{Faculty of Physics, University of Vienna, Boltzmanngasse 5, Vienna A-1090, Austria}

\author{C. Meingast}
%\email{christoph.meingast@kit.edu}
\affiliation{Institut für Festkörperphysik, Karlsruhe Institute of Technology, 76021 Karlsruhe, Germany}

\maketitle
\section{Supplemental material}
\subsection{Measurement technique}
To measure $Y_{[110]}$, platelet-like single crystals of K-Ba122 and Co-Ba122 were cleaved and cut to a size of $L\times b\times h\sim(2-3\times1\times0.1)\textnormal{ mm}^3$ with $h$ along the crystalline $c$ axis and $L$ along tetragonal [110]. The crystals were supported by three wires and then placed in the capacitance dilatometer\cite{Meingast1990} (see Fig. 1 (a,b) of the main article), in which they naturally experience a small and roughly constant force of about $0.2$ N due to the springs holding the upper capacitor plate. This force is normally used to hold the sample in place; here it acts to bend the sample slightly, and the temperature-dependent changes of this 'bending-modulus' are directly obtained by monitoring the degree of bending using the separation of the capacitor as a gauge. In order to obtain quantitative results, the force applied by the dilatometer springs was calibrated, the initial degree of bending was determined, and the results were corrected for thermal expansion of the cell and the sample. More details of this procedure will be reported separately\cite{Böhmerunpublished}. Besides being able to measure small and thin samples, this technique yields very high resolution data due to the sensitivity of the capacitance method. 

\subsection{Calculation of $Y_{[110]}$}
For thin samples ($h/L\ll1$) in the proper orientation (see Fig. 1 (b) of the main article), the bending spring constant $k_s$ is related to $Y_{[110]}$, the Young's modulus (\textit{i. e.} the modulus of elasticity for uniaxial tension) along the [110] direction\cite{Kityk1996}: $k_s\approx4b(h/L)^3Y_{[110]}$. 
$Y_i$ is the inverse of $S_{ii}$, the relevant component of the elastic compliance tensor\cite{Kityk1996} so that for a tetragonal sample
\begin{equation}
Y_{[110]}=4\left(\frac{1}{C_{66}}+\frac{1}{\gamma}\right)^{-1}\textnormal{; } \gamma=\frac{C_{11}}{2}+\frac{C_{12}}{2}-\frac{C_{13}^2}{C_{33}}\label{eq:2}\\
\end{equation}
which means that $Y_{[110]}$ is dominated by $C_{66}$ as long as $C_{66}$ is smaller than $\gamma$.

\subsection{Coupling between nematic susceptibility and superconductivity}

\begin{figure}
\includegraphics[width=8.6cm]{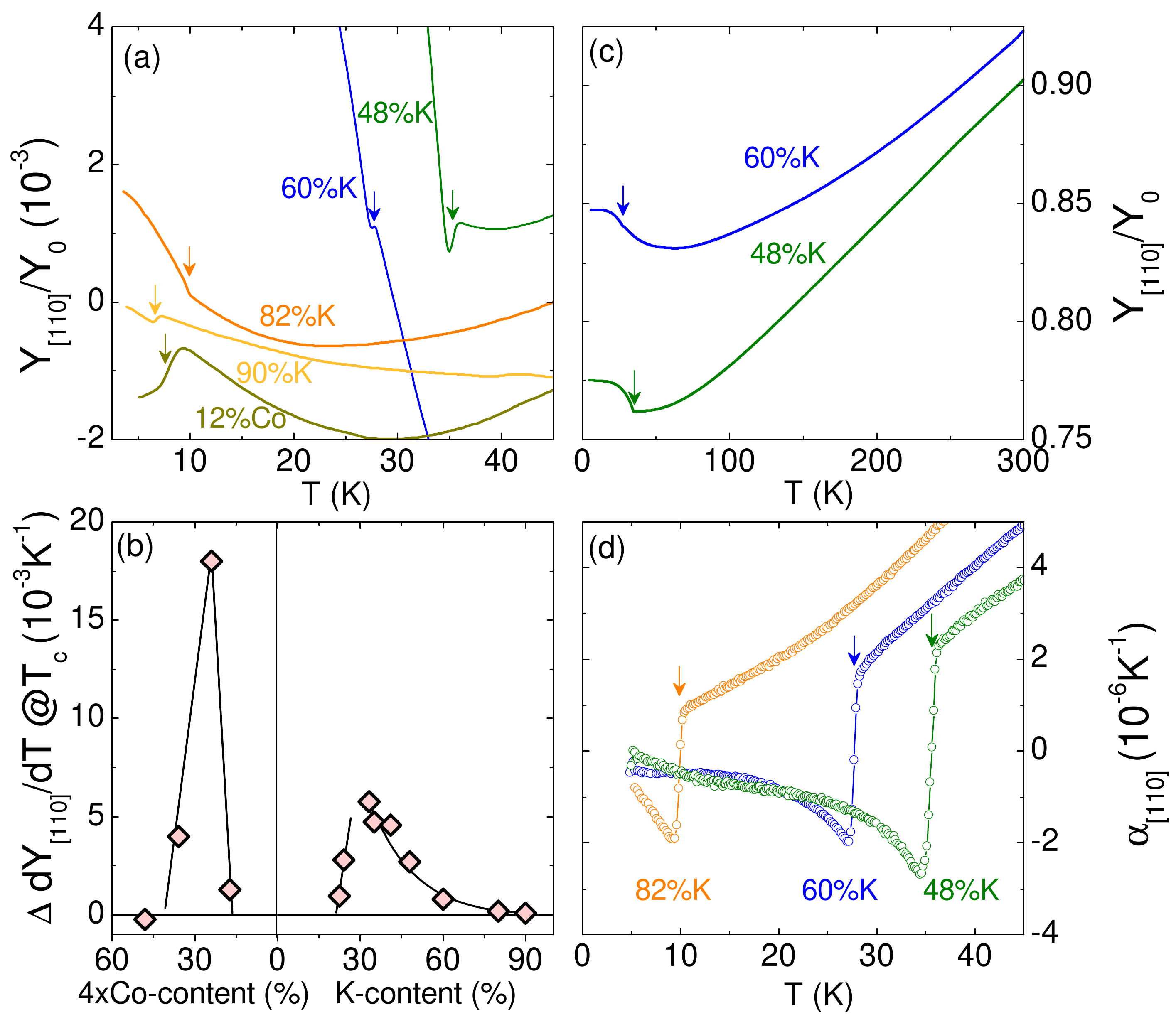}
\caption{(color online). (a) Young's modulus $Y_{[110]}/Y_0$ of overdoped samples close to $T_c$ on a magnified scale. Data have been vertically offset by arbitrary values. (b) Impact of superconductivity on the nematic susceptibility as quantified by the discontinuous change in the slope $dY/dT$ at $T_c$. (c) $Y_{[110]}/Y_0$ of K-Ba122 samples with intermediate doping levels. (d) Bulk superconducting transitions of the same samples seen in the in-plane thermal-expansion coefficient $\alpha_{[110]}$. Arrows in (a), (c), and (d) indicate $T_c$.}
\label{fig:4}
\end{figure}

The high resolution of our data allows to resolve even the small effects at high doping levels which occur close to $T_c$ (see Fig. \ref{fig:4}(a)). Near optimal doping, $Y_{[110]}$ hardens significantly as a response to superconductivity, in agreement with previous reports\cite{Fernandes2010, Goto2011, Yoshizawa2012}. We observe an additional, small step-like softening of the Young's modulus at $T_c$ at intermediate doping levels ($x=48-60$ \% K), which is surprisingly absent at $82\%$ K. Highly overdoped samples ($x = 90$ \% K and $x = 12$ \% Co) exhibit only the step-like anomaly, which might however mask a tiny hardening below $T_c$, if existent. This step-like softening of $Y(T)$ at $T_c$ is actually the usual behavior expected from thermodynamics at a second order phase transition and is related to the stress-derivative of $T_c$, \textit{i.e.} the normal coupling between superconductivity and the lattice\cite{Hardy2009}. 

The hardening, on the other hand, directly reflects the competition between superconductivity and the nematic order parameter $\varphi$ affecting the lattice via equation 2 of the main article \cite{Fernandes2010, Yoshizawa2012II}. In Fig. \ref{fig:4}(b) the effect of this competition, as quantified by the discontinuous slope change of $Y/Y_0(T)$ at $T_c$ (ignoring the step-like anomaly), is plotted versus doping. For both Co-Ba122 and K-Ba122 the slope change is maximal near optimal doping and then decreases strongly for overdoped samples with a weak tail extending up to at least 82\% K on the hole-doped side. The maximum value is however three times larger for Co-Ba122 than for K-Ba122, suggesting that the coupling of the nematic order to superconductivity is significantly stronger in this system. Finally, we point out that \textit{e.g.} the 60 \% K sample exhibits a broad hardening of $Y_{[110]}$ starting at $\sim60$ K , which is clearly unrelated to the onset of superconductivity. To prove this, we show in Fig. \ref{fig:4}(d) the bulk transitions of these samples as determined by thermal expansion, which are all quite sharp. 

\end{document}